\documentclass[aps,prl,twocolumn,groupedaddress,showpacs]{revtex4}
\usepackage{amsmath}
\usepackage{amsfonts}
\begin{document}

\title{Influence of Absorbers on the Electromagnetic Radiation}
\author{Neil V. Budko}
\affiliation{Laboratory of Electromagnetic Research, Faculty of Electrical
Engineering, Mathematics and Computer Science,
Delft University of Technology,
Mekelweg 4, 2628 CD Delft, The Netherlands}
\email{n.v.budko@tudelft.nl}

\date{\today}

\begin{abstract}
The phenomenon of the electromagnetic absorption by arbitrarilly distributed
discrete absorbers is analyzed from the photon point of view.
It is shown that apart from the decrease in the intensity of the signal the net
effect of absorption includes a relative increase in the photon bunching.
\end{abstract}

\pacs{42.50.-p, 42.50.Nn}

\maketitle

When photons get absorbed nothing particularly interesting seems to happen
from either classical or quantum points of view.
If there is an absorber somewhere
in the way of the electromagnetic radiation, it will disturb the initial
wave. This disturbance will then spread around at the spead of
light and may be detected with the corresponding time delay.
This is the classical causal picture. The photon picture is quite different.
Absorption of a {\it single} photon
does not have a classical analogue. The classical electromagnetic field
decays gradually in time and never disappears all at once.
Whereas, the absorbed photon is instanteneously and completely removed from
the field, even if this photon was the whole field.
The process is rather trivial from the theoretical point of view though. The
event of absorption is described by the annihilation operator
sending the photon to the vacuum state, which does not intefere with the
other states of this photon. We are not talking about
the general vacuum field here, which does produce nontrivial effects.

The practical modeling of losses is often very simplistic
even in single photon counting situations. Normally, a macroscopic
absorption coefficient is used to account for the
reduced intensity of the signal \cite{Brassard}. In experiments on quantum
optics one tries to minimize or compensate
any losses as they destroy the sub-poissonian
statistics and other low-intensity quantum effects \cite{Kiss}. It is
believed that optical losses are equivalent to a random sampling of
an already quite random stream of photons, so that the resulting stream is
even more random \cite{Fox}.
In other words, most of what we know about absorption is either bad or
useless.
Yet there is definitely something strange about it as well.
For example, we might ask ourselves a question:
Does the absorption of a photon at one side of an omnidirectional source
instantaneously change the field at the other side?
More generally: What influence do absorbers have on the electromagnetic
field, in particular, on the visibility of the source?
This Letter gives some partial answers to these questions.
Perhaps not so surprisingly,
it turns out that even non-obscuring absorbers
negatively influence the visibility of the radiating source.
A little less obvious is the higher survival probability of the bunched
photons
as compared to the stream of separate photons.

Consider an electromagnetic source with a given vacuum probablility $P$ to
detect a photon at a given location.
For simplicity we shall think of an omnidirectional point source, so that
this probability is approximately
 \begin{equation}
 \label{eq:VacProb}
 P\sim\frac{1}{4\pi r^{2}},
 \end{equation}
and is the same for all observation angles.
Let there be a (point) absorber present at a distance $r_{1}$ from the
source, able to absorb one photon at a time. As we are interested in the
specific
effect of absorption, we shall consider an absorber which does not re-emit
the photon, but completely removes it from the radiated field.
What is the probability to detect a single photon at some point on a sphere
with radius $r_{2}>r_{1}$? Clearly, it is not equal to the
vacuum probability (\ref{eq:VacProb}) as the photon may be absorbed by the
first absorber and never even reach the sphere with radius $r_{2}$.
It is also clear that due to the noninterfering nature of the absorbed
state,
the classical probability calculus should be applied to account for the lost
photons.

Let us denote by $P(n)$ the probability for the photon to be absorbed
(detected) by the $n$-th absorber (detector) situated at the sphere
with radius $r_{n}$. In general, it is given by
 \begin{equation}
 \label{eq:VacProbR1}
 P(n)=P(r_{n})P(n\vert r_{n}),
 \end{equation}
where $P(r_{n})$ is the probability for a photon to reach the sphere and
$P(n\vert r_{n})$ is the conditional probability for a photon
to be detected at the $n$-th detector situated at some point on this sphere
provided it has reached $r_{n}$.
Since there is vacuum between the first absorber and the source, and
$P(r_{1})=1$,
the probability $P(1)$ is well-defined, i.e.,
 \begin{equation}
 \label{eq:VacProbR12}
 P(1)=P(1\vert r_{1})\sim\frac{1}{4\pi r_{1}^{2}}.
 \end{equation}
Conditional probabilities $P(n\vert r_{n})$ are just the vacuum
probabilities, whereas $P(r_{n})$ depend on the presence
of absorbers at distances smaller than $r_{n}$ from the source,
irrespectively of the position of these absorbers on the
surfaces of their corresponding spheres (for a general source -- on the
equipotential surfaces of the probability density function).
For example, with just one absorber at $r_{1}$ the probability to reach
$r_{2}$ is
 \begin{equation}
 \label{eq:ProbR2}
 P(r_{2})=1-P(1).
 \end{equation}
Hence,
 \begin{equation}
 \label{eq:Prob2}
 P(2)=\left[1-P(1)\right]P(2\vert r_{2}).
 \end{equation}
If there are $N-1$ absorbers present at arbitrary locations on the spheres
with radii $r_{n}$, $r_{n+1}>r_{n}$, $n=1,\dots N$,
then the probability to detect a single emitted photon at the $N$-th
location at distance $r_{N}$ from the source, $N>1$, is given by
 \begin{equation}
 \label{eq:ProbN}
 P(N)=\left[1-\sum\limits_{n=1}^{N-1}P(n)\right]P(N\vert r_{N}),
 \end{equation}
which may be viewed as a recurrent formula starting with $P(1)$ defined in
(\ref{eq:VacProbR1}). Thus we
conclude that even non-obscuring absorbers have a negative influence on the
visibility of the omnidirectional source, i.e.
 \begin{equation}
 \label{eq:ProbNvsProbRnN}
 P(N)<P(N\vert r_{N})=P_{\rm vacuum}.
 \end{equation}

Electromagnetic sources consist of many atoms which emit arbitrary,
generally uncorrelated, streams of photons.
To distinguish this case from a single photon emission we shall add an extra
argument in our probabilities
denoting the number of photons where necessary. Consider a fixed time
interval $T$ during which
the source emits a
stream of $K$
mutually noninterfering photons coming one at a time.
For an omnidirectional source and independent emission events the
probability to detect all $K$ photons during $T$ (plus retardation due to
travel time) is
 \begin{eqnarray}
 \label{eq:ProbKN}
P(K,N)&=&\left[P(N)\right]^{K}
\\ \nonumber
&=&\left[1-\sum\limits_{n=1}^{N-1}P(n)\right]^{K}\left[P(N\vert
r_{N})\right]^{K},
 \end{eqnarray}
which is again smaller than the vacuum probability $[P(N\vert r_{N})]^{K}$.
Probability to detect $M$ photons from the stream of $K$ photons is given by
 \begin{eqnarray}
 \label{eq:ProbMN}
 &P(M,N)&=C^{K}_{M}\left[P(N)\right]^{M}\left[1-P(N)\right]^{K-M}
 \\ \nonumber
&=&\frac{K!}{M!(K-M)!}\left[1-\sum\limits_{n=1}^{N-1}P(n)\right]^{M}\left[P(
N\vert r_{N})\right]^{M}
 \\ \nonumber
 &\times&\left[1-\left[1-\sum\limits_{n=1}^{N-1}P(n)\right]P(N\vert
r_{N})\right]^{K-M}.
 \end{eqnarray}

Photons may also come in bunches. For example, $K$ photons may be
emitted as a single bunch at some arbitrary moment within $T$.
If $K>N$, then at least one photon will always reach the $r_{N}$-sphere as
each absorber can absorb only one photon at a time, letting all the rest
through.
Thus, for $K>N$ the probability to detect at least
one photon at the $N$-th detector is equal to one.
If $K-N=M$, then at least $M$ photons will always reach $r_{N}$ during the
retarded time interval $T$
and with a suitable (multiphoton) detector we can detect all of them with
the probability
 \begin{equation}
 \label{eq:ProbKNbunched}
 P'(M,N)>\left[P(N\vert r_{N})\right]^{M}.
 \end{equation}
One can show that
 \begin{eqnarray}
 \label{eq:Compare}
 P_{\rm separate} < P_{\rm bunched}<P_{\rm vacuum},
 \end{eqnarray}
which is easy to understand, if we simply compare the probabilities to reach some
given sphere. Obviously, in the $K>N$ case, 
$P_{\rm bunched}(K-N,r_{N})=1$, whereas, $P_{\rm separate}(M,r_{N})<1$ for any $M$.

The situation with photons and absorbers reminds a herd of gazelles trying
to cross a river with a crocodile in it.
A crocodile can only catch one animal at a time. Hence, crossing the river
together increases the survival probability
of the herd as a whole. Similarly, atoms and individual charged particles of
the source may emit
photons in arbitrary completely uncorrelated sequences. However, when
crossing the space full of absorbers,
it is mostly the bunched photons that survive. Thus, apart from the usual decrease of the signal intensity, 
the net effect of absorption is to increase whatever bunching tendencies the original field had. Certainly, things get more
complicated when not only absorption,
but also re-emission (i.e. scattering) is involved. The natural first guess
would be to use
the quantum probabilities accounting for the lossless coherent and partially
incoherent
scattering instead of the simple vacuum probabilities used above. In that
case
our calculations apply only to those photons that are absorbed irreversibly.
It is not clear at the moment whether the increase in the survival probability of the bunched photons will
persist when the scattering effects are taken into account.


\end{document}